\documentclass[11pt]{article}
\usepackage{epsfig}
\usepackage{citesort}
\usepackage{times}

\newcommand{\kin}[1]{\sqrt{2 E_{#1} \over m_e}}
\newcommand{\kinup}[1]{\sqrt{2 E^{#1} \over m_e}}
\newcommand{\re}[1]{(\ref{#1})}
\newcommand{\eq}{\begin{equation}}
\newcommand{\eqx}{\end{equation}}
\newcommand{\eqn}{\begin{eqnarray}}
\newcommand{\eqnx}{\end{eqnarray}}

\newcommand{\nin}{\noindent}

\begin{document}
\nin

\begin{center}
{\Large \bf Multi-electron-recombination rates estimated within dense plasmas}\\
\vspace{10mm}

{\large B.~Ziaja $^{\ast,\dag}$ \footnote{Author to whom correspondence should be addressed. Electronic mail: ziaja@mail.desy.de}, 
 F.~Wang $^{\ast}$, and E.~Weckert $^{\ast}$}

\vspace{3mm}
$^{\ast}$ \it Hamburger Synchrotronstrahlungslabor, Deutsches Elektronen-Synchrotron, Notkestr. 85, D-22603 Hamburg, Germany\\

\vspace{3mm}
$^{\dag}$ \it Department of Theoretical Physics,
 Institute of Nuclear Physics,
\it Radzikowskiego 152, 31-342 Cracow, Poland\\
 
\end{center}

\vspace{10mm}
\centerline{\Large October 2008}

\nin
{\bf Abstract:}
{\em We investigate the rates for multielectron recombination within a dense plasma environment in local thermodynamic equilibrium (LTE). We find that these multielectron recombination rates can be high within dense plasmas, and they should be treated in the simulations of the plasmas created by intense radiation, in particular for plasmas created by intense VUV radiation from free-electron-laser (FEL) or for modelling the inertial confinement fusion 
(ICF) plasmas.    
}\\ 


\section{Introduction}

Electronic many-body recombination processes that occur within dense, cold plasmas have attracted much attention recently as a possible explanation of the strong energy absorption within laser-created plasmas \cite{brabec,rost1}.
During the first cluster experiment performed at the free-electron-laser FLASH facility at DESY \cite{desy,desy2} such plasmas were created after the irradiation of the xenon clusters with VUV photons of energy, $E_{\gamma}=12.7$ eV. Pulse duration did not exceed $50$ fs, and the maximal pulse intensity was, $I\leq 10^{14}$ W/cm$^2$. Highly charged Xe ions (up to $+8$) of high kinetic energies were detected, indicating a strong energy absorption that could not be explained using the standard approaches \cite{desy5,desy2,desy7}. More specifically, the energy absorbed was almost an order of magnitude larger than that one predicted with classical absorption models, and the ion charge states were much higher than those observed during the irradiation of isolated atoms at the similar conditions. This indicated that at these radiation wavelengths some processes specific to many-body systems were responsible for the enhanced energy absorption. 

The lowest order many-body recombination is three-body recombination, where two continuum electrons are involved initially, and in the final state one of these electrons is captured by the target ion. The excess energy released by the recombining electron is then carried away by the other outgoing electron so that the three-body recombination does not involve any emission of photons \cite{hahn}. For this to happen it is necessary that the electron density must be high enough so that the collisional processes are more probable than the dissipative radiation processes \cite{eliezer}. The three-body recombination is the inverse of the electron collisional ionization, and this relation is used in deriving the relevant cross-section formula, by detailed balance.

In the similar way higher-order recombination processes can be described. More than two electrons are then involved. Rates for these processes strongly depend on the electron density, and can again be calculated with the rates of the corresponding inverse many-body collisional processes.

The mechanism of plasma heating due to many-body recombination that was proposed in Ref.\ \cite{brabec} in order to explain the experimental results of Ref. \cite{desy} can be best described as a sequence of Auger processes followed by photoionization. According to \cite{brabec}, the strongly coupled nanoplasma is created shortly after the exposure. Within this plasma, a probability of finding two or more electrons being close to an ion is relatively high. One electron can be captured by the ion, while the other electrons (spectators) carry away the excess energy released by the recombining electron. The captured electron may then absorb an radiation photon and become ionized once more. This cycle repeats many times during the duration of the pulse, leading to an efficient heating of the electron cloud. Ions of charge up to $+7$ were predicted with this model for the $Xe_{1000}$ cluster.

Here we aim to study a different many body recombination process that may also 
occur within dense plasmas. This is multielectron recombination (MER) 
that is an inverse process to multiple collisional ionization. Several electrons then recombine simultaneously, and a single spectator electron carries away the excess energy. We will evaluate the significance of this process within plasmas created by VUV FEL radiation from xenon and argon clusters \cite{desy} and also within inertial-confinement-fusion plasmas \cite{dicklee}. We will first derive analytical formulas for MER cross-section and rates, using the microscopic reciprocity relation and the detailed balance principle \cite{oxenius}.
Using these formulas, we will obtain numerically the MER rates for Xe and Ar plasmas. These rates will give an estimate, how significantly these processes 
contribute to the dynamics within dense plasmas.

\section{Calculation of rates for multielectron recombination}

\begin{figure}
\vspace*{0.5cm}
\centerline{(a) \epsfig{width=6cm, file=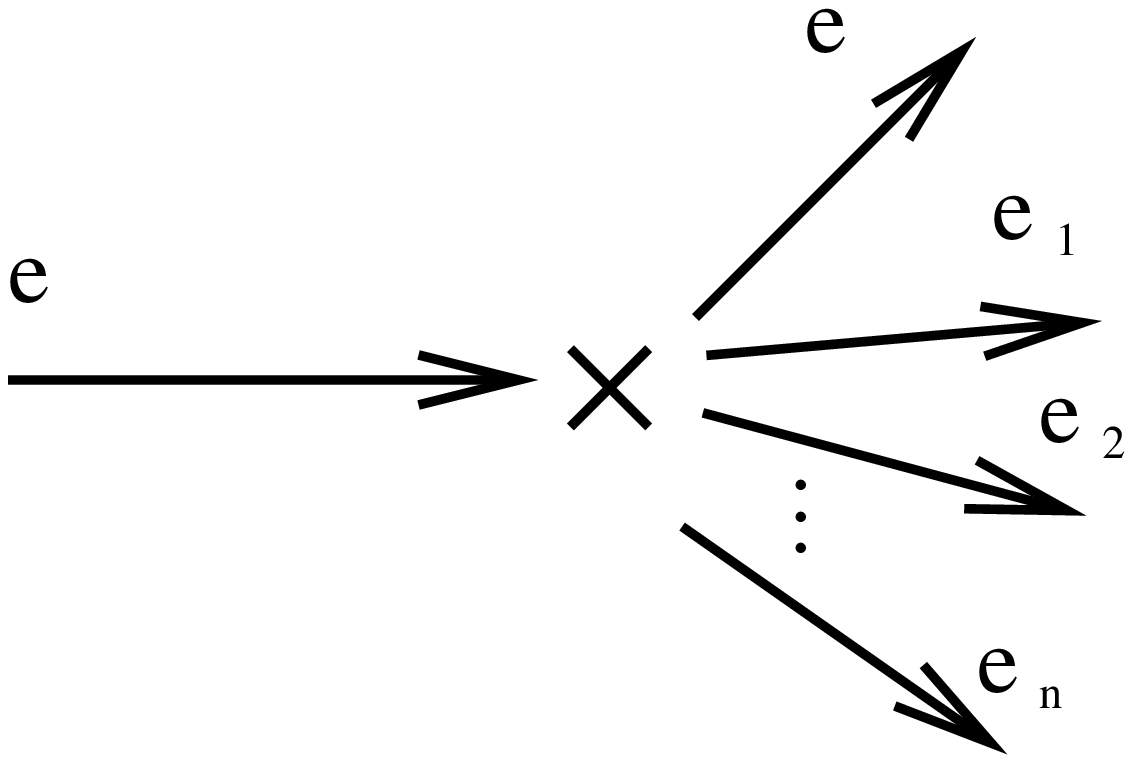}
(b) \epsfig{width=6cm, file=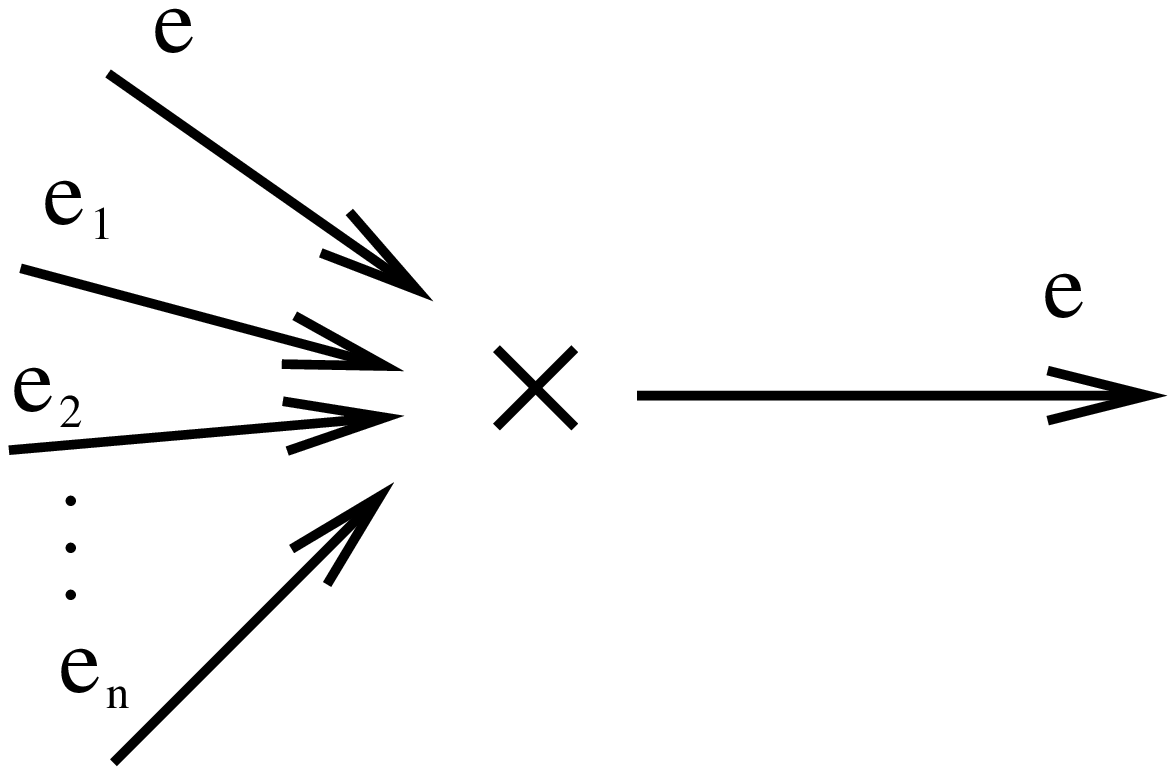}}

\caption{Collisional processes within a dense plasma: (a) multiple collisional ionization and (b) its inverse process: multielectron recombination. The electron, $e$, is a spectator.}
\label{f1}
\end{figure}

MER is an inverse of the multiple collisional process (Fig. \ref{f1}), where
the primary electron ionizes an atom, releasing one or more secondary electrons. At $n=1$ we arrive at the three-body recombination (one-electron recombination), at $n=k$ we have k-electron recombination. 
In what follows we will use the notation as in Ref.\ \cite{oxenius}.

Derivation of the MER cross sections and rates bases on the validity of the quantum mechanical reciprocity relation:
\eq
w(i\rightarrow f)=w(f \rightarrow i),
\label{reci}
\eqx
where $w(i\rightarrow f)$ is the transition probability per unit time for the transition $|i> \rightarrow |f>$ between single quantum states of the states
$i$ and $f$. Reciprocity relation \re{reci} is an essential condition for the validity of the principle of detailed balance \cite{oxenius}. The conditions
for validity of the reciprocity relation can be obtained within the framework of the general scattering matrix theory \cite{oxenius}. The result is that the reciprocity relation $w(i\rightarrow f)$  is not generally valid, not even if the Hamiltionian of the system is invariant under space reflection (P) and time reversal (T). However, relation \re{reci} holds in the framework of perturbation theory, and thus applies to radiative processes and to collisions in the Born
approximation. The reciprocity relation holds moreover for all collision processes involving unpolarized particles, provided that the Hamiltonian is 
P- and T-invariant. In what follows we will consider only unpolarized particles, therefore the reciprocity relation will be valid for the processes considered.

Explicit form of the reciprocity relation for specific processes can be derived
directly from quantum mechanics. Here, following Ref. \cite{oxenius}, we will
derive it through the principle of detailed balance, by writing down the rate equations and using the explicit form of the thermal distribution functions. 
This procedure is simple and straightforward, while the relations so found are
independent of this method, and are identical with those derived directly from 
quantum mechanics. We will assume the electron distributions to be isotropic.

Reaction equation for a collisional process ($k$-body multiple ionization) and its inverse process ($k$-body multielectron recombination) in Fig.\ \ref{f1} can be written as:
\eq
A(E_0)+e(E)\rightleftharpoons A^{+}(E_+)+e(E^{\prime})+e(E_1)+\ldots+e(E_k),
\label{rea}
\eqx
where $A$ denotes atom or ion in the state of initial energy $E_0$, $A^+$ denotes the ionized atom or ion in the state of energy $E^+$, $e$ is electron 
of energy $E(E_i)$. Energy of the spectator electron changes from $E\rightleftharpoons E^{\prime}$ during the process. This reaction leads to the balanced rate equation:
{\footnotesize
\eqn
\rho_0\,\kin{}\, \rho_e\,f_e(E)\,\Omega_{0+}(E;E^{\prime},E_1,\ldots,E_k)dE^{\prime}
dE_1\ldots dE_k=\rule{3cm}{0cm}&&\nonumber\\
\rho_+\,\kinup{^{\prime}}\,\rho_e\,f_e(E^{\prime})\,dE^{\prime}
\kin{1}\,\rho_e\,f_e(E_1)\,dE_1\ldots
\kin{n}\,\rho_e\,f_e(E_k)\,dE_k
\Omega_{+0}(E^{\prime},E_1,\ldots,E_k;E)\,dE,&&
\label{rate}           
\eqnx
}
where we introduced the following notation: $\rho_0$ denotes density of atoms/ions in initial energy state $E_0$, $\rho_+$ denotes density of atoms/ions in the final energy state $E_+$, $\rho_e$ is the electron density and $m_e$ is the electron mass. The distribution, $f_e(E)$ is a normalized isotropic Maxwell-Boltzmann distribution of electron kinetic energy, and quantities $\Omega_{0+}(E;E^{\prime},E_1,\ldots,E_n)$ and $\Omega_{+0}(E^{\prime},E_1,\ldots,E_n;E)$ are differential cross sections (in respect to energy) for the multiple collisional ionization and multielectron recombination respectively. The angular distribution of electrons was assumed
to be isotropic, and the integration over scattering angles was performed.

The rate for multielectron recombination is defined as:
{\footnotesize
\eqn
R_{k(RC)}&=&\rho_e^{k+1}\, \int\, \kin{^{\prime}}\kin{1}\ldots\kin{k}
f_e(E^{\prime})\,f_e(E_1)\ldots\,f_e(E_k)\,
\Omega_{+0}(E^{\prime},E_1,\ldots,E_k;E)\nonumber\\
&&\rule{1cm}{0cm}dE^{\prime}\,dE_1\,\ldots\,dE_k\, dE,
\label{rcrate}
\eqnx
}
and can be related to the collisional multiionization rate:
\eq
R_{k(I)}=\rho_e\, \int\, \kin{}\,f_e(E)\,
\Omega_{0+}(E;E^{\prime},E_1,\ldots,E_k)\,
dE^{\prime}\,dE_1\,\ldots\,dE_k\, dE,
\label{irate}
\eqx
through a relation that follows from Eq.\ \re{rate}:
\eq
\rho_+\,R_{k(RC)}=\rho_0\,R_{k(I)}.
\label{frate}
\eqx

In a balanced state the relation between the atom/ion distributions in states
$E_0$ and $E_+$ is given by a Saha relation \cite{oxenius} known also as principle of detailed balance. For k-multielectron recombination and 
k-multiple ionization this relation can be derived (not shown here) from Eq.\ \re{rea} as:
\eq
{\rho_0 \over \rho_+}={g_0 \over g_+}\, e^{\beta\,E_{+0}}\,
\left( {1\over 2} \rho_e\,\lambda_e^3 \right)^k,
\label{saha}
\eqx
where $\beta=1/kT$ is the inverse of the temperature within the system and $\lambda_e={h \over {\sqrt{2 \pi m_e kT} }}$ is the thermal de Broglie wavelength of free electrons. Coefficients, $g_0,g_+$, are statistical weights of the bound energy levels, $E_0, E_+$, respectively, and $E_{+0}=E_+-E_0$ is the energy 
needed for transition from state $E_0$ to state $E_+$ (ionization energy). 
After substituting Eq.\ \re{saha} to Eq.\ \re{frate}, we obtain:
\eq
R_{k(RC)}=R_{k(I)}\,{g_0 \over g_+}\,e^{\beta\,E_{+0}}\,
\left( {1\over 2} \rho_e\,\lambda_e^3 \right)^k.
\label{ffrate}
\eqx

From Eq.\ \re{rate} also a microscopic relation between ionization and recombination cross sections can be obtained:
\eq
g_0\,E\, \Omega_{0+}(E;E^{\prime},E_1,\ldots,E_k)=
g_+{2^{4k}\over h^{3k} }\,(\pi\,m_e)^k\,E^{\prime}\,E_1\,\ldots\,E_k\,\,
\Omega_{+0}(E^{\prime},E_1,\ldots,E_k;E).
\label{micro}
\eqx
This is a generalization of the Fowler relation obtained for three-body recombination \cite{oxenius,lee}. As expected, at $k=1$ Eq.\ \re{micro} reduces to the Fowler relation:
\eq
g_0\,E\ \Omega_{0+}(E;E^{\prime},E_1)=
g_+{2^{4}\over h^{3} }\,(\pi\,m_e)\,E^{\prime}\,E_1\,
\Omega_{+0}(E^{\prime},E_1;E).
\eqx

For plasmas in a non-equilibrium state (non-LTE plasmas) Eq.\ \re{ffrate} is not longer valid. The non-equilibrium rate for multielectron recombination can then be still obtained from Eq. \re{rcrate}, using the recombination cross section
derived from the microscopic relation, Eq. \re{micro}, and convoluted with non-equilibrium electron distributions. The difficulty lies in the correct parametrization of the differential multiple ionization cross section, which
is not known for an arbitrary value of $n>1$ \cite{rbeb0,rbeb}.
Therefore we restrict here to LTE plasmas, for which the direct rate equation, Eq.\ \re{ffrate}, can be used.

\section{Estimation of multielectron recombination rates for xenon and argon}

We used Eq.\ \re{ffrate} to estimate the rate for multielectron recombination within plasmas in thermal equilibrium. In order to calculate the collisional ionization rate, we used experimentally determined total cross sections for collisional multiionization \cite{tawara}.
 
\begin{figure}
\vspace*{0.5cm}
(a)\epsfig{width=7cm, file=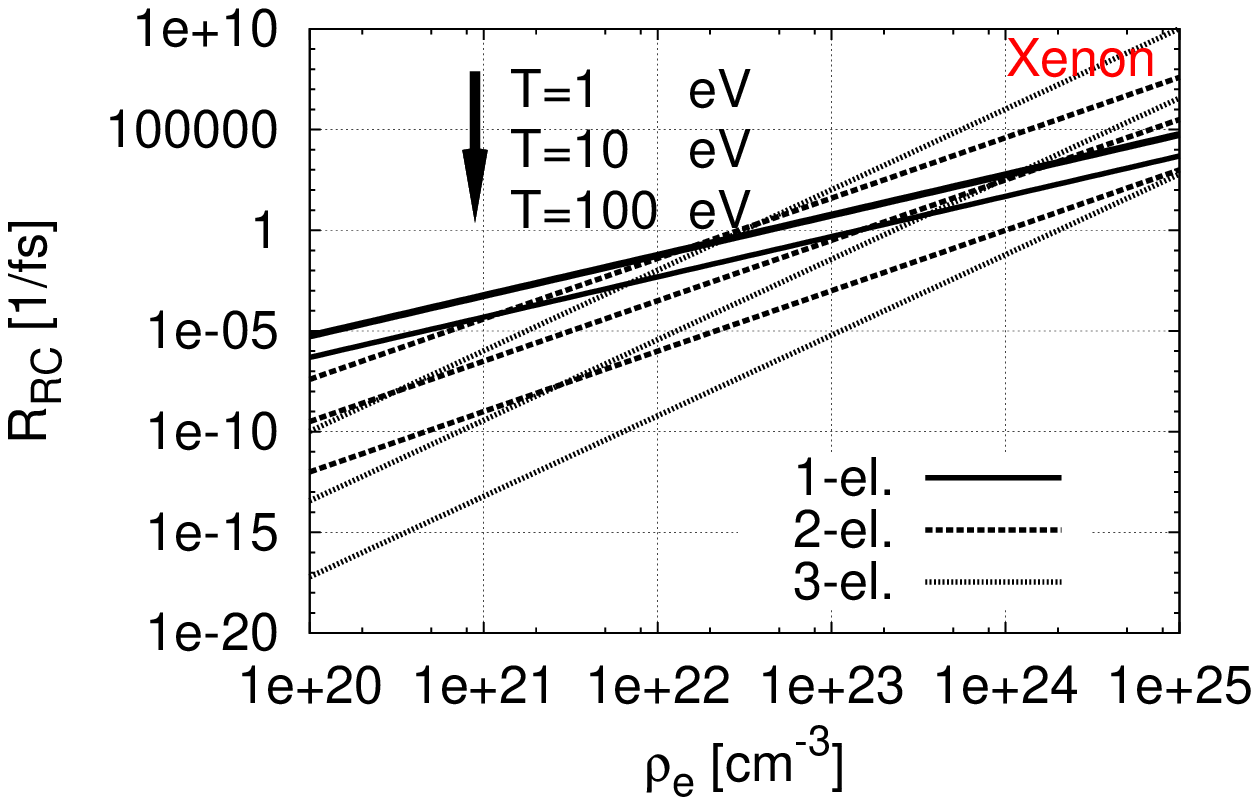}
(b)\epsfig{width=7cm, file=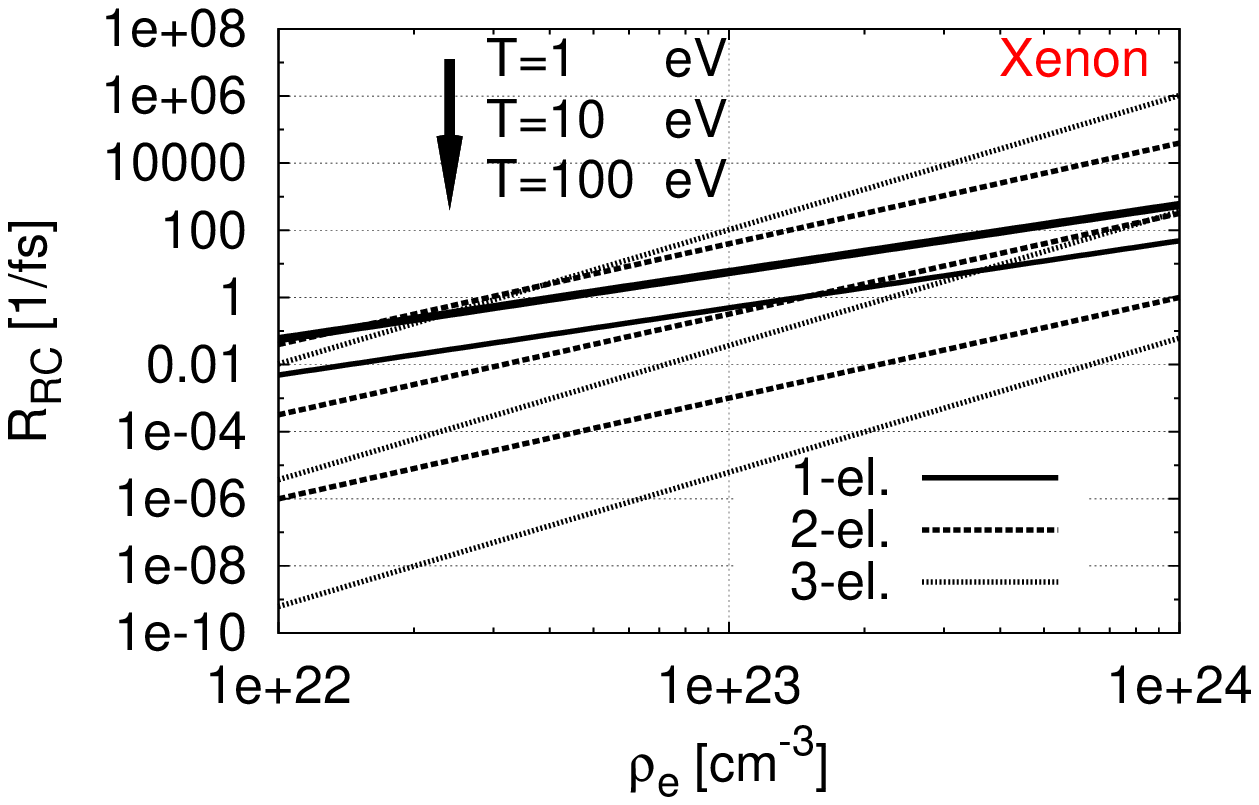}\\
(c)\epsfig{width=7cm, file=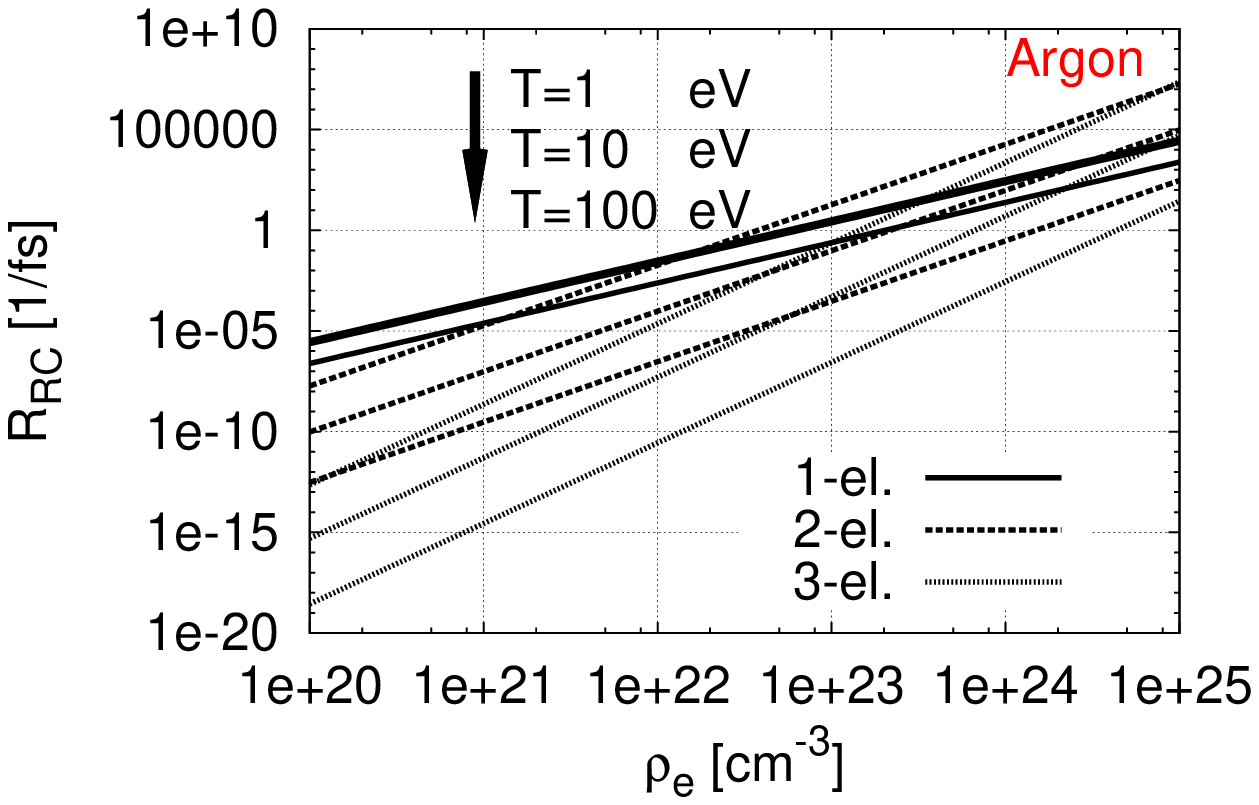}
(d)\epsfig{width=7cm, file=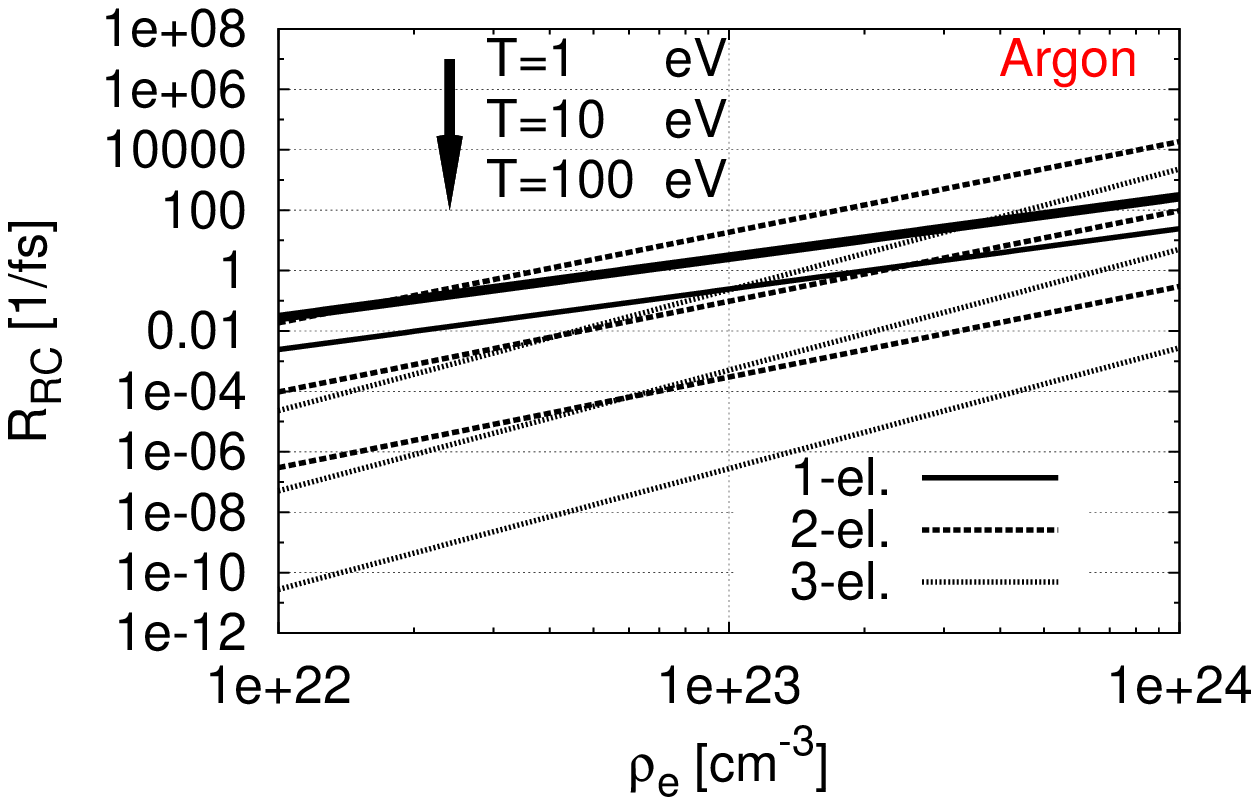}
\caption{Rates for multielectron recombination: $1,2,3$-electron recombination to average Rydberg state as functions of plasma density for plasma in LTE. They are plotted for: (a) xenon with a closeup (b), and (c) argon with a closeup (d). Arrows show how the rates change with increasing values of electron temperature, $T=1,10,100$ eV.}
\label{rydberg}
\end{figure}

Eq.\ \re{ffrate} implies that the electrons recombine in most cases
to the higher Rydberg states, for which the statistical weights, $g_0$, are  large. Therefore here we estimate the recombination rates to an "average" Rydberg state. This is the state closest to the average of the available states within the plasma. This average is obtained with weights that correspond to the
statistical weights, $g_0$, of the considered states. The recombination rates obtained are plotted in Fig.\ \ref{rydberg}. Density range, 
$\rho=10^{22}-10^{24}$ cm$^{-3}$ at temperatures $T=1-10$ eV corresponds to plasmas created by intense VUV laser pulse from FEL \cite{ziajab2,ziajab3,ziajab4,ziajab5}. The rates obtained for densities $\rho=10^{22}-10^{25}$ cm$^{-3}$ and temperatures $T \sim 100$ eV are relevant 
for inertial confinement fusion plasmas.

At the density, $\rho=10^{23}$ cm$^{-3}$, and temperatures, $T=1-10$ eV, the highest rates are: the rate for 3-electron recombination (Xe), and 2-electron recombination (Ar). This tendency keeps with increasing electron density. 
The absolute values of recombination rates ($n=1-3$) at $\rho=10^{23}$ cm$^{-3}$ are then between $10^{-2}-10^2$ $1/$fs for Xe and $10^{-4}-10^1$ $1/$fs for Ar, depending on the temperature.

In the region of high densities, $\rho >10^{24}$ cm$^{-3}$, and at high temperatures, $T=100$ eV, relevant for ICF plasmas, the absolute recombination rates are high: $10^3-10^4$ $1/$fs for Xe, and $10^1-10^3$ $1/$fs for Ar at $\rho=10^{25}$ cm$^{-3}$.
The 1-electron recombination is predominant but the other recombination rates
($n=2,3$) are less by the factor of 10 (Xe) and 10-100 (Ar).  

\section{Summary}

The aim of our study was to estimate the multielectron recombination rates ($n>1$) within dense plasmas.

We derived the MER recombination rate for LTE plasmas built of xenon or argon atoms, using the experimental cross sections for multiple ionization of atoms. We considered the recombination to the average Rydberg state. Plasma effects as lowering of ionization thresholds, modifications of the cross sections have not been treated. Cut-off of available atomic states due to the plasma environment was included.

We showed that in the $\rho-T$ region relevant for plasmas created after 
irradiation of Xe and Ar clusters with intense VUV pulses from FEL, MER recombination dominates over the 1-electron recombination. The values of recombination rates ($n=1-3$)
are between $10^{-2}-10^2$ $1/$fs for Xe and $10^{-4}-10^1$ $1/$fs for Ar at $\rho=10^{23}$ cm$^{-3}$, depending on the temperature. In the $\rho-T$ region of ICF plasmas, the absolute values of MER are high ($10^3-10^4$ $1/$fs for Xe, and $10^1-10^3$ $1/$fs for Ar at $\rho=10^{25}$ cm$^{-3}$). The 1-electron recombination dominates in this regime but the 2- and 3-electron recombination rates are less by factor of 10-100. 

Our estimations imply that the MER rates should be treated in simulations of processes occurring within ICF plasmas and FEL laser created plasmas.
The quantitative validation how significantly the MER processes contribute to the plasma dynamics would require a detailed rate equation approach, where both competing processes: non-sequential ionization and MER recombination are included, and plasma effects have to be treated. This approach is, however, beyond the scope of the present study.
 
\section*{Acknowledgments}

Beata Ziaja is grateful to R. W. Lee, H.-K. Chung, S. Toleikis and T. Tschentscher for illuminating discussions and comments. 

\end{document}